\documentclass{nature}

\usepackage{graphicx}
\usepackage{amsmath}
\usepackage{amssymb}
\usepackage{textcomp}
\usepackage{gensymb}
\usepackage{xcolor}
\usepackage{lineno}

\newcommand{\reftextit}[1]{}
\newcommand{\vv}[1]{\mathbf{#1}}
\newcommand{\rr}[1]{\mathrm{#1}}
\newcommand{\mat}[1]{\bar{\mathrm{#1}}}
\newcommand{\TODO}[1]{#1}
\newcommand{\TODOO}[1]{#1}

\makeatletter
\let\saved@includegraphics\includegraphics
\AtBeginDocument{\let\includegraphics\saved@includegraphics}
\renewenvironment*{figure}{\@float{figure}}{\end@float}
\makeatother

\title{Phonon Renormalization in Reconstructed MoS$_\text{2}$ Moir\'e Superlattices}

\author{Jiamin Quan$^{1,\dagger}$, Lukas Linhart$^{1,2,\dagger}$, Miao-Ling Lin$^{3}$, Daehun Lee$^{1}$,
    Jihang Zhu$^{1}$, Chun-Yuan Wang$^{1}$, Wei-Ting Hsu$^{1}$, Junho Choi$^{1}$, Jacob Embley$^{1}$, Carter Young$^{1}$, Takashi Taniguchi$^{4}$, Kenji Watanabe$^{4}$, Chih-Kang Shih$^{1}$, Keji Lai$^{1}$, Allan H. MacDonald$^{1}$, Ping-Heng Tan$^{3,*}$, Florian Libisch$^{2,*}$\&Xiaoqin Li$^{1,*}$}
    
\begin{document}
\maketitle

{\renewcommand{\baselinestretch}{1.5}
\begin{affiliations}
    \item Department of Physics and Center for Complex Quantum Systems, 
    The University of Texas at Austin, Austin, TX, USA.
    \item Institute for Theoretical Physics,
    Vienna University of Technology, 1040 Vienna, Austria, EU.
    \item State Key Laboratory of Superlattices and Microstructures, Institute of Semiconductors, Chinese Academy of Sciences, 100083 Beijing, China.
    \item National Institute for Material Science, 1-1 Namiki, Tsukuba, Ibaraki 305-0044, Japan.\\
    \thanks{$^{\dagger}$ These authors contributed equally to this work.}\\
    $^{*}$\thanks{Corresponding authors:\textcolor{blue}{{elaineli@physics.utexas.edu};  {florian.libisch@tuwien.ac.at;}\\{phtan@semi.ac.cn}}}
\end{affiliations}
}
{\renewcommand{\baselinestretch}{1.7}
\begin{abstract}
In moir\'e crystals formed by stacking van der Waals (vdW) materials, surprisingly diverse correlated electronic phases and optical properties can be realized by a subtle change in the twist angle. Here, we discover that phonon spectra are also renormalized in MoS$_2$ twisted bilayers, adding a new perspective to moir\'e physics. Over a range of small twist angles, the phonon spectra evolve rapidly due to ultra-strong coupling between different phonon modes and atomic reconstructions of the moir\'e pattern. We develop a new low-energy continuum model for phonons that overcomes the outstanding challenge of calculating properties of large moir\'e supercells and successfully captures essential experimental observations. \TODO{Remarkably, simple optical spectroscopy experiments can provide information on strain and lattice distortions in moir\'e crystals with nanometer-size supercells.} The newly developed theory promotes a comprehensive and unified understanding of structural, optical, and electronic properties of moir\'e superlattices.
\end{abstract}
}
\maketitle
\renewcommand{\baselinestretch}{2}
\setlength{\parskip}{7pt}

In vertical van der Waals (vdW) homo- or heterobilayers with weak interlayer coupling, a finite twist angle between layers leads to a moir\'e superlattice that induces periodic modulations of atomic structure, energy, and optical selection rules\cite{zhang2017interlayer,tran2019evidence}. Controlling the twist angle with $\sim$ 0.05-0.1~$^\circ$ accuracy \cite{kim2016van} in graphene bilayers near the magic angles leads to completely different correlated electronic phases including superconductivity\cite{cao2018unconventional,yankowitz2019tuning}, orbital magnetism\cite{sharpe2019emergent,serlin2020magnetic}, and correlated insulator states\cite{cao2018correlated}. Similar phenomena have been observed in transition metal dichalcogenide (TMD) twisted bilayers (TBLs)\cite{regan2019optical,tang2019wse2,shimazaki2020strongly}, although with a reduced sensitivity to the twist angle\cite{naik2018ultraflatbands,wang2019magic}. \TODO{In WSe$_2$ TBLs, correlated insulating states are observed over a broad range of twisted angles between 4$^\circ$ and 5.1$^\circ$, while superconductivity only emerges at $\theta=5.1^\circ$, indicating intriguing changes over a magic twist angle continuum\cite{wang2019magic}.}

These prior experiments on moir\'e superlattices have been interpreted using a rigid lattice model 
in which the local atomic stacking is assumed to be determined by rotating pristine two-dimensional lattices. 
However, theoretical studies and microscopy experiments have shown that substantial lattice relaxation can occur in TMD TBLs\cite{gadelha2020lattice,carr2018relaxation,sushko2019high,weston2019atomic,mcgilly2019seeing}. \TODO{Recent piezoresponse force microscopy (PFM) and scanning transmission electron microscopy (STEM)\cite{mcgilly2019seeing,weston2019atomic} measurements reveal a tessellated pattern of mirror-reflected triangular 3R domains in twisted TMD homobilayers, separated by a network of thin domain boundaries for twist angles $\theta<$2$^\circ$. }\TODOO{This detailed structural information challenges the interpretations of previous experiments based on rigid lattice picture.} 

In this work, we investigate how atomic reconstructions influence  Raman spectra in MoS$_2$ TBLs with twist angles $\theta$ ranging from 0$^\circ$ to 20$^\circ$. \TODO{The accurate twist control (0.1$^\circ$) allows us to discover a systematic renormalization of phonon spectra not addressed in previous works\cite{huang2016low,puretzky2016twisted,holler2020low}.} Guided by computation, Raman experiments can be used to identify three regimes of the moir\'e patterns: the relaxed regime (0$^\circ \leq \theta <$2$^\circ$), the transition regime (2$^\circ \leq \theta<$6$^\circ$), and the rigid lattice regime ($\theta \ge$6$^\circ$). In the relaxed and rigid regimes, the Raman spectra hardly change with twist angle. In the transition regime, however, low frequency interlayer shear (S) and layer-breathing (LB) modes evolve rapidly with the twist angle. We further identify a splitting of the commonly observed high frequency intralayer E$_{2g}$ mode that is attributed to the distortion of the hexagonal lattice within each monolayer.
\TODO{The rapid phonon spectra evolution over a small range of twist angles cannot be explained by any existing model, e.g., simple phonon dispersion folding in moir\'e crystals\cite{lin2018moire} or linear chain models\cite{tan2012shear}.}
To explain these observations, we develop a simple low-energy continuum model to describe the phonon modes of TBLs at any small twist angle in the presence of lattice reconstructions, a challenging task for atomistic models due to the large supercell size and required supercell commensurability. The excellent agreement between experiment and theory validates our model and enables unambiguous spectral assignments.  Furthermore, the theoretical approach allows us to describe phonons and electrons on an equal footing, paving the way for a more comprehensive understanding of electron-phonon coupling in moir\'e superlattices. 

The atomic reconstruction of the moir\'e pattern is determined by a twist angle dependent competition between strain and interlayer coupling\cite{jung2015origin,carr2018relaxation,rosenberger2019atomic,weston2019atomic} as shown in Fig.~1 (See Supplementary I for detailed calculation). At the smallest twist angles, the real space supercell is very large, allowing substantial lattice relaxation even though it is driven by weak van der Waals interactions between the layers and inhibited by strong in-plane bonding within each layer. The relaxation pattern forms large triangular regions in which the energetically favorable stacking configurations (see Fig.~1c, right column and Supplementary I), known as AB(BA) stacking (or 3R stacking for $\theta=0^\circ$, see Supplementary II) are approached\cite{zhang2018structural,carr2018relaxation}. The local strain (left column in Fig.~1c) in the relaxed regime peaks along domain boundaries and at topological defects \cite{alden2013strain} with AA stacking where domains intersect.

As the twist angle increases and real space supercell size decreases, the distance between neighboring AB and BA stacking configurations is reduced (Fig.~1d). Correspondingly, the area occupied by the domain walls that interpolate between them increases steadily across a transition regime. Finally, the TBL reaches the rigid regime at large twist angles ($\theta\ge$6$^\circ$). In this regime (Fig.~1e), the area with nearly perfect low-energy and high-symmetry AB(BA) stacking is small. The resulting reduction of strain leads to essentially flat\cite{huang2016low} rigid layers \cite{nam2017lattice,carr2018relaxation}. The evolution of the low-energy AB(BA) area (red line) and the high-energy AA (green line) area as a function of twist angle is summarized in Fig.~1f (top), while the area-averaged strain is summarized in Fig.~1f (bottom). \TODO{These variations of atomic configuration and local strain are expected to modulate the lattice vibrations.}

We measured Raman spectra from a series of MoS$_2$ TBLs with accurately controlled twist angles in a range of angles  0$^\circ \leq  \theta \leq  20^\circ$. Details of the sample preparation procedure and Raman measurements are discussed in  the methods section and the Supplementary III. The measured spectra feature several phonon modes divided into the low-frequency (Fig.~2a) and high-frequency range (Fig.~2b). The low-frequency Raman spectra exhibit two types of phonon modes, the interlayer S and LB modes in which the relative motion of the two monolayers is  parallel or perpendicular to the two-dimensional (2D) layers, respectively \cite{molinasanchez2011bilayermos2,tan2012shear,zhang2013raman,zhao2013interlayer} (Fig.~2a top). The LB Raman modes have a fine structure due to coupling with discrete LB modes of the hexagonal Boron Nitride (hBN) substrate with a finite thickness\cite{lin2019cross} (see Supplementary V). We remove this fine structure\cite{lin2019cross} (Supplementary Fig.~S6) to focus on the main features related to moir\'e pattern reconstruction via a fast Fourier transform filter (see Supplementary VI and VII). As the twist angle increases, one branch of the LB mode (LB$_1$) blueshifts and seems to disappear along with the S mode, while a second branch of the LB mode (LB$_2$) emerges. There are two dominant intralayer modes in the high-frequency Raman spectra (Fig.~2b), commonly denoted as E$_{2g}$ (~385 cm$^{-1}$) and A$_{1g}$ (~407 cm$^{-1}$) following the assignments appropriate for the D$_{\rr{6h}}$ symmetry of bulk (2H stacked) MoS$_{2}$ \cite{lee2010anomalous,verble1970}. The two-fold degenerate E$_{2g}$ mode originates from opposite motions of two sulfur atoms relative to the Mo atom within the 2D plane while the A$_{1g}$ mode arises from the out-of-plane relative vibrations of the sulfur atoms\cite{li2012bulk,lee2010anomalous} (Fig.~2b top). Although the A$_{1g}$ mode frequency is nearly independent of twist angle, the E$_{2g}$ mode evolves into a doublet in the 2$^\circ\leq$$\theta$$<$6$^\circ$ transition regime which we will discuss in more detail below. \TODO{While Raman on TMD TBLs has previously been reported\cite{huang2016low,puretzky2016twisted,holler2020low,liao2020precise,debnath2020evolution}, those experiments were performed on samples with much less control of the twist angle and missed the systematic phonon renormalization captured by our experiments.} 

The evolution of the Raman spectra is further analyzed by tracking the peak positions and linewidths as a function of twist angle (Figs.~3a-c). Distinct features emerge in the three regimes. In the relaxed regime (0$^\circ\leq$$\theta$$<$2$^\circ$), the frequencies and linewidths of all modes exhibit little change because the moir\'e patterns remain qualitatively the same (matching the $\theta=0^\circ$ case\cite{molinasanchez2011bilayermos2}) with only quantitative changes in the AB(BA) domain area. In the transition regime (2$^\circ\leq$$\theta$$<$6$^\circ$), the phonon spectra change drastically. The S mode broadens and quickly disappears. There is also a rapid and systematic change in the frequency, linewidth, and intensity (see Supplementary IX).  Prominently, the evolution of the LB modes resembles an anti-crossing behavior (Fig.~3a) typically observed when hybrid modes form due to coupling between different phonon modes, as we explain below. The linewidth of each phonon mode directly reflects the phonon lifetime. Assuming\cite{cusco2016temperature} that the linewidth $\gamma_{\rr{FWHM}}$ is related to the lifetime $\tau$ by $\ \tau=\hbar/\gamma_{\rr{FWHM}}$, we find that the lifetime of the LB$_1$ mode is 49 ps and 7.1 ps for TBLs at 2$^\circ$ and 3.3$^\circ$, respectively. \TODO{These drastic changes in phonon frequency and lifetime in the transition regime may be used to infer twist angle and supercell size on the basis of the TBL Raman spectra alone, offering a simple and powerful spectroscopy technique to characterize moir\'e crystals. The onset of the transition regime identified by Raman spectra at $\theta\sim2^\circ$ agrees remarkably well with a very recent STEM study performed on mechanically stacked MoS$_2$ TBLs\cite{weston2019atomic}.} Finally, in the rigid regime with $\theta\ge$6$^\circ$, incommensurate stacking again results in stable Raman spectra with little dependence on the twisted angle. In this regime, the exponential dependence of the polarizability on the layer separation substantially reduces the intensity of the S mode with a substantially reduced intensity, explaining why it is not observed in our experiment\cite{huang2016low, puretzky2016twisted, holler2020low} (see also Supplementary X).

\TODO{Clear signatures of three reconstruction regimes are also found in the high frequency E$_{2g}$ mode (Fig.~3c). In the relaxed regime, strain is absent except on the sharp domain walls and strain is also small in the rigid regime. Thus, no clear peak splitting is observed in either of these two regimes. By contrast, the mode splits into a doublet E$_{2g}^{+}$ and E$_{2g}^{-}$ in the transition regime due to local strain caused by the atomic reconstruction. High strain locally distorts the hexagonal unit cell and breaks the three-fold rotational symmetry ultimately causing the E$_g$ mode to split \cite{lee2017strain,wang2013raman} (Fig.~3d). The largest splitting of the E$_{2g}$ mode, up to 3.2 cm$^{-1}$, occurs at $\theta =$ 2.5$^{\circ}$ TBL where the influence of strain on hexagonal symmetry is maximal (Fig.~1f bottom). Our measurements report averaged strain under the laser spot. Local variations of strain within a moir\'e supercell can be resolved using a near-field technique \cite{gadelha2020lattice}. However, reaching a sufficient spatial resolution in the transition regime where the supercell size ranges from 9 nm (2 $^\circ$) to 3 nm (6 $^\circ$) would be very challenging.}

\TODO{We confirm the presence of atomic reconstructions in a TBL at $\theta\sim0.08^\circ$ by PFM measurements. The large strain gradients near the AA stacking regions and the domain walls (Figs.~1c-e) allow piezoelectric coupling to an out-of-plane a.c. electric field\cite{mcgilly2019seeing} (see Methods). PFM data in Fig.~3e clearly reveals the reconstructed moir\'e superlattice with a typical size of $\sim230$ nm as expected. The superlattice is divided into large triangular AB/BA domains by narrow domain walls that locally break the single-layer D$_{3h}$ symmetry (Fig.~3d). The results confirm the high quality of our samples and the expected moir\'e reconstructions in the relaxed regime.}

\TODO{Simple models such as phonon dispersion folding\cite{lin2018moire} or effective force constant models\cite{koshino2019_phonon,tan2012shear,maity2019} cannot explain the complicated evolution of modes we observe. Full ab-initio calculations become too expensive at small twist angles\cite{huang2016low} and are, like effective force constant models, limited to commensurate twist angles (Fig.~4a).} Here, we adapt a low-energy continuum model approach, first developed by Bistritzer and MacDonald\cite{jung2014ab,bistritzer2011moire} to calculate the electronic system of moir\'e superlattices, to phonons in TBLs (See Supplementary XI for details).  The pristine lattice vectors $\vv a_1=(a_0,0), \vv a_2=a_0 (-1,\sqrt{3})/2$ (see Fig.~4b) define the adjoint reciprocal lattice vectors $\vv G$.  In the small angle limit, the relationship between the displacement $\vv d$ between layers (which characterizes local stacking) and position within the moir\'e pattern, maps $\vv G$ onto the reciprocal lattice vectors $\tilde{\vv G}$  of the moir\'e cell via $\tilde{\vv{G}}(\theta,\vv G) \approx -\theta \hat{z} \times \vv{G}$ (see Fig.~4c). We calculate phonon modes from local crystalline dynamical matrices $\mat{D}$ evaluated as a function of the displacement $\vv d$ (Fig.~4b). To calculate the optically active phonon modes near the central $\Gamma$ point, we assemble the moir\'e dynamical matrix $\mat{D}_{\rr{m}}(\vv q,\vv q')$ from matrices $\mat{D}{(\vv{q}|\vv{d})}$ evaluated for each local stacking $\vv d$ via Fourier transform,
\begin{linenomath*}
\begin{equation}
    \mat{D}_{\rr{m}}(\vv{q},\vv{q}') = \sum_{\vv{G,d}} 
        \delta \big(\vv{q} - \vv{q}' - \tilde{\vv{G}}(\theta,\vv G)\big)\, \mat{D}(\vv{q}'|\vv{d})e^{i\vv d\cdot\vv G}.
        \label{eq:bmdforphonons}
\end{equation}
\end{linenomath*}
This moir\'e dynamical matrix is off-diagonal in reciprocal space because of the slow spatial variation of $\vv d$. We find that small $|\vv{G}|$ terms dominate in Eq.~\ref{eq:bmdforphonons} allowing us to truncate the sum after the first shell of six non-zero reciprocal lattice vectors, coupling each $\vv{q}$-point in the moir\'e reciprocal unit cell to six replicas. In particular, the central $\Gamma$-point is coupled to six neighbors $\tilde{\vv{G}}_{1\ldots 6}$ (blue dots in Fig.~4c). The $\mat{D}(\vv{q}|\vv{d})$ for each displacement $\vv d$ can be straightforwardly obtained from density functional perturbation theory. This model does not include any free parameters. 

\TODO{When truncating the expansion after the first shell our model yields a total of 126 modes, i.e. 18 modes that are folded by the moir\'e reciprocal lattice and evolve continuously with twist angle.}  Mode energies calculated at the $\Gamma$ point of MoS$_2$ probed by Raman spectroscopy neglecting (including) lattice relaxation are shown in Fig.~4d (e). We rescale the overall interlayer coupling strength by a single factor of 1.15 to match the interlayer DFT frequencies at $\theta=0$ to the experiment (see Supplementary XI).

The LB modes are only weakly twist-angle dependent in both the relaxed (LB$_1$ $\approx 40$ cm$^{-1}$) and the rigid (LB$_2$ $\approx 33$ cm$^{-1}$) regimes (Fig.~4e). Calculations and measurements match perfectly for the LB modes at all $\theta$. We find a prominent anti-crossing of the LB modes caused by coupling to the dispersive folded transverse acoustic (TA) modes, in excellent agreement with the experimental evolution of the LB modes in the transition regime (Fig.~3a). At $\theta \approx 3.5^\circ$, the folded TA acoustic phonon modes originating at $\tilde{ \vv G}=\tilde{\vv G}_1\ldots \tilde{\vv G}_6$ (gray lines in Fig.~4d) are degenerate with the LB mode at $\vv{q} = \Gamma$. The ultra-strong coupling between the two modes can be explained by the large $\vv d$-dependence of $\mat D(\vv q'|\vv d)$ in the corresponding spacial direction. By comparing calculations without (Fig.~4d) and with (Fig.~4e) lattice relaxations, one recognizes that the LB modes are similar in both cases. Thus, although the rapid evolution of the LB modes coincides with the transition regime, our calculations suggest this is not caused by the atomic reconstructions of the moir\'e pattern.

\TODO{Only when accounting for relaxation of the lattice via elasticity theory (Fig.~4e and Supplementary I and XI C), the instability (e.g. imaginary frequency for the S mode) of the rigid TBLs for $\theta\lesssim 4^{\circ}$ is removed. At small twist angles, lattice reconstruction causes the degenerate S mode  frequencies to shift and match the measured values.} For larger angles, our model overestimates the corrugation (variation of layer separation in $\vv d$) of the moir\'e pattern\cite{huang2016low}, causing deviations between the calculated and measured S mode frequencies (see Supplementary X). Neglecting its atomistic structure, the moir\'e patterns preserve $C_{3}$ symmetry even in the presence of lattice relaxation and therefore conserve the degeneracy of the S mode at the $\Gamma$-point present for $\theta=0^\circ$. Given the complex interplay of layer separation, mode coupling and lattice reconstruction, a detailed explanation for the behavior of the linewidth is outside the scope of the present work. 

We also calculate the evolution of the A$_{1g}$ and E$_{2g}$ Raman peaks with twist angle. While the optically active A$_{1g}$ mode is hardly affected by the moir\'e, local strain lifts the degeneracy of the E$_{2g}$ mode by breaking the hexagonal symmetry\cite{wang2013raman,lee2017strain}. The non-uniform strain present in moir\'e structures (Figs.~1c-e) becomes uniaxial at the domain walls (Fig.~3d), breaking the single-layer D$_{3h}$ symmetry and causing splitting of the E$_{2g}$ mode into E$_{2g}^{\pm}$ \cite{lee2017strain}. The observed proportionality between the splitting E$_{2g}^+ - $E$_{2g}^-$ (see Fig.~4f) and the average strain (see Fig.~1f) further corroborates our model and underpins the crucial role of strain in these systems. 

In summary, our experiments have revealed that phonon spectra are renormalized in MoS$_2$ moir\'e supperlattices. \TODO{Three regimes of atomic reconstructions can be identified by distinct Raman spectra. The most interesting Raman spectral changes in the transition regime suggest a continuous and subtle evolution of atomic configurations and strain. Such information is partially accessible via scanning tunneling microscopy\cite{gadelha2020lattice} but challenging for many common scanning probe and near-field techniques. In the same regime, a rich variety of electronic phases in TMD TBLs have been reported\cite{wang2019magic}}, \TODOO{highlighting the importance of reconciling electronic phases with static and dynamic lattice properties.} To treat the phononic and electronic degrees of freedom on an equal footing, we introduce a computationally efficient method to describe phonons in moir\'e crystals at any small twist angle. In the future, one can extend this model to describe electron-phonon interaction which may be critical to explain superconductivity in moir\'e superlattices\cite{cao2018unconventional,cao2018correlated,wang2019magic}.

\newpage
\begin{methods}
\textbf{Sample preparation} \newline
 The samples were fabricated using a modified “tear-and-stack” technique\cite{kim2016van}. A schematic diagram of the stacking process is presented in Fig.~S3a. hBN and monolayer MoS$_{2}$ flakes were mechanically exfoliated from a bulk crystal onto a polydimethylsiloxane sheet. The hBN thickness is typically around 15 nm, measured by atomic force microscopy. The MoS$_{2}$ monolayer was identified by optical contrast and Raman spectroscopy. The bottom hBN was first transferred onto the Si/SiO$_{2}$ (90 nm) substrate and subsequently annealed  at 500 ºC for 10 hours. The van der Waals force between hBN and MoS$_{2}$ monolayer was used to tear a part of the monolayer flake at room temperature, which was transferred onto the hBN. The separated monolayer pieces were rotated by a specific angle and stacked together. The accuracy in controlling the twist angle is $\sim$ 0.1$^\circ$. Finally, the samples were annealed under ultrahigh vacuum (around 10$^{-5}$ mbar) at 150 ºC for 2 hours to enhance the coupling between two layers. We carefully checked the microscope images before and after annealing and found no rotations during the annealing for all samples (see Supplementary Figs.~S3b,c).\TODO{The spatial uniformity of optical properties of the twisted bialyers was futher confirmed via low-frequency Raman mapping (see Supplementary Fig.~S4).}\par
 
\noindent
\textbf{Raman measurement} \newline
Raman spectra were measured at room temperature using a Princeton Acton 7500i spectrometer equipped with a liquid-nitrogen-cooled charge-coupled detector (CCD), a $\times$100 objective lens (numerical aperture = 0.90). BragGrate notch filters were used to reject the Rayleigh scattering down to 8 cm$^{-1}$. The excitation laser was a 532 nm (i.e. 2.33 eV) continuous-wave laser from a Verdi V10. A 1200 lines/mm grating was used in the Raman measurements. An incident laser power of 0.1 mW is used to avoid sample heating. The excitation laser and collected Raman signal are collinearly polarized.

\noindent
\textbf{Piezoresponse force microscopy measurements.}\newline
In a PFM measurement, an a.c. bias is applied on the conductive tip to induce sample deformation through the piezoelectric effect. The amplitude and phase of vertical (out-of-plane) and horizontal (in-plane) deformations of the sample are recorded during the contact-mode scan, providing local information of the electromechanical response. The experiments were performed on a commercial atomic-force microscope (XE-70 AFM, Park Systems). A lock-in amplifier (HF2LI, Zurich instruments) was used to apply the a.c. bias (typically around 1 V) and demodulate the PFM signals. The radius of the cantilever probe (ANSCM-PT-10, App Nano Inc.) is less than 30 nm and the force constant is 1$\sim$5 N/m. For out-of-plane PFM, the 1st harmonic of the cantilever resonance (80 kHz $\sim$ 90 kHz) was used for detection. For in-plane PFM, the 3rd harmonic frequency (320 kHz $\sim$ 330 kHz) was used for detection.

\noindent
\textbf{Density functional perturbation theory (DFPT) calculations} \newline
To build the low energy continuum model, two types of DFPT calculations are required. For the $\vv{q}$-dependent parts of the moir\'e dynamical matrix we calculate one single layer of MoS$_2$ using a $6\times6\times1$ super cell. For the $\vv{q}$ independent part, we employ 10$\times$10 individual primitive cell bilayer MoS$_2$ DFPT calculations sampling the different stackings $\vv d$. We use VASP with a $\vv k$-point Monkhorst-Pack grid of $17\times17\times1$ (3$\times$3$\times$1 for the super cell) with an energy cut-off of 400 eV and a unit cell height of 35 $\textrm{\AA}$ to provide sufficient vacuum between the layers\cite{kresse1993vasp,kresse1994vasp}. For simplicity we use the local-density-approximation (LDA) (for further details see Supplementary XII).

\noindent
\textbf{Calculation of strain and lattice relaxation}\newline
Strain was calculated by minimizing the total energy functional \begin{linenomath*}\[\rr{U}_{\rr{tot}} = \rr{U}_{\rr B}[\vv{u}^{t},\vv{u}^{b}] + \rr{U}_{\rr E}[\vv{u}^{t}] + \rr{U}_{\rr E}[\vv{u}^{b}],\]\end{linenomath*} with respect to the displacements $\vv u^{t(b)}(\vv r)$. We model the potential energy functional $\rr U_{\rr B}$ via a generalized stacking fault energy (GSFE)\cite{carr2018relaxation,nam2017lattice} obtained from sampling the configuration space using DFPT. The 2D elastic energy functional $\rr U_{\rr E}[\vv{u}^{t(b)}]$ is modeled using the Lam\'e parameters $\lambda=3.29$ eV/$\textrm{\AA}^{2}$ and $\mu = 3.6$ eV/$\textrm{\AA}^{2}$ \cite{carr2018relaxation} (for further details see Supplementary I).

\noindent

\end{methods}

\begin{addendum}
    \item The spectroscopy experiments at UT-Austin (J.Q.) were primarily funded by the U.S. Department of Energy, Office of Basic Energy Sciences under grant DE-SC0019398 and a grant from the University of Texas. Material preparation was funded by the Welch Foundation via grant F-1662. The collaboration between the X.L., C.S., K.L., and M.A. groups is facilitated by the NSF-MRSEC under DMR-1720595, which funded J.C. and J.E. partially. L.L. and F.L. acknowledge support by the TU-D doctoral program of TU Wien, as well as from the Austrian Science Fund (FWF), project I-3827. The authors acknowledge discussions with S. Reichardt and the use of facilities and instrumentation supported by the National Science Foundation through the Center for Dynamics and Control of Materials: an NSF MRSEC under Cooperative Agreement No. DMR-1720595. P.T. acknowledges support from  the National Natural Science Foundation of China (Grant No.11874350) and CAS Key Research Program of Frontier Sciences (Grant No. ZDBS-LY-SLH004). M.L. acknowledge the support from the Project funded by China Postdoctoral Science Foundation (Grant No. 2019TQ0317). The PFM work (D.L. and K.L.) was supported by NSF DMR-2004536 and Welch Foundation Grant F-1814. K.W. and T.T. acknowledge support from the Elemental Strategy Initiative conducted by the MEXT, Japan, Grant Number JPMXP0112101001, JSPS KAKENHI Grant Numbers JP20H00354 and the CREST(JPMJCR15F3), JST.
    \item[Author contributions] J.Q. led the optical experiments, M.L., C.W., W.H. J.E. and J.C. assisted the experiment. L. L. led the theoretical calculations. D. L. performed PFM measurements. J.Q. and C.Y. prepared the twisted bilayer sample. T.T. and K.W. provided the hBN sample. J.Q., L.L., F.L. and X.L. wrote the manuscript. X.L., F.L., A.H.M., P.T., K. L., and C.S. supervised the project. All authors contributed to discussions and  writing the manuscript.

\end{addendum}
\textbf{References} \\
\setlength{\baselineskip}{15pt}
\setlength{\parskip}{7pt}
\bibliography{ref}

\begin{figure}
    \centering
   \includegraphics[width=1\textwidth]{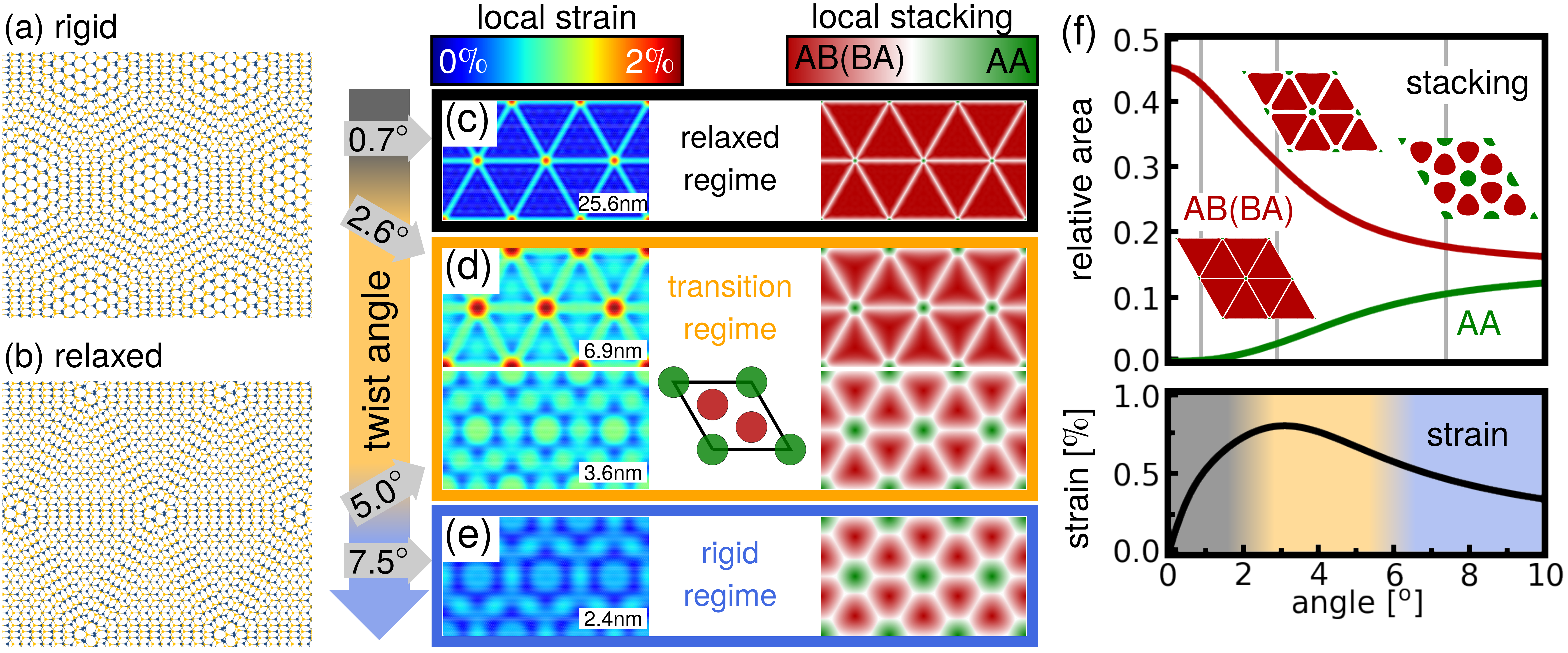}
    \caption{Twist angle dependent lattice reconstruction in  MoS$_2$ TBLs with small twist angles.
    (a)-(b) Two rotated layers of MoS$_2$
    (a) without and (b) with lattice relaxation. 
    (c)-(e) Calculated patterns of local strain (left column) and stacking (right column) at various twist angles. 
    The plots are drawn in moir\'e cell units to facilitate comparison of quasi periodic super cells of different sizes.
    Three distinct lattice reconstruction regimes can be identified:(c) the relaxed regime (black rectangle), 
     (d) the transition regime (orange rectangle) and (e) the rigid regime (blue rectangle).
    The inset illustrates the area of AA (dark green) and AB(BA) stacking (dark red) within a rigidly rotated bilayer.
    (f) (top) Fraction of the total area covered by AB(BA) (dark red) or AA stacking (dark green).   Area is counted as AA/AB(BA) if the relative displacement between the layers is smaller than 0.25 times the lattice constant.
    The insets represent three representative lattice stackings corresponding to twist angles at 0.7$^{\circ}$, 2.6$^{\circ}$ and 7.5$^{\circ}$.
    (bottom) Evolution of the average (local) strain in the system (exact definition as in Supplementary I).}
    \label{fig:fig1}
\end{figure}

\begin{figure}
   \centering
   \includegraphics[width=0.6\textwidth]{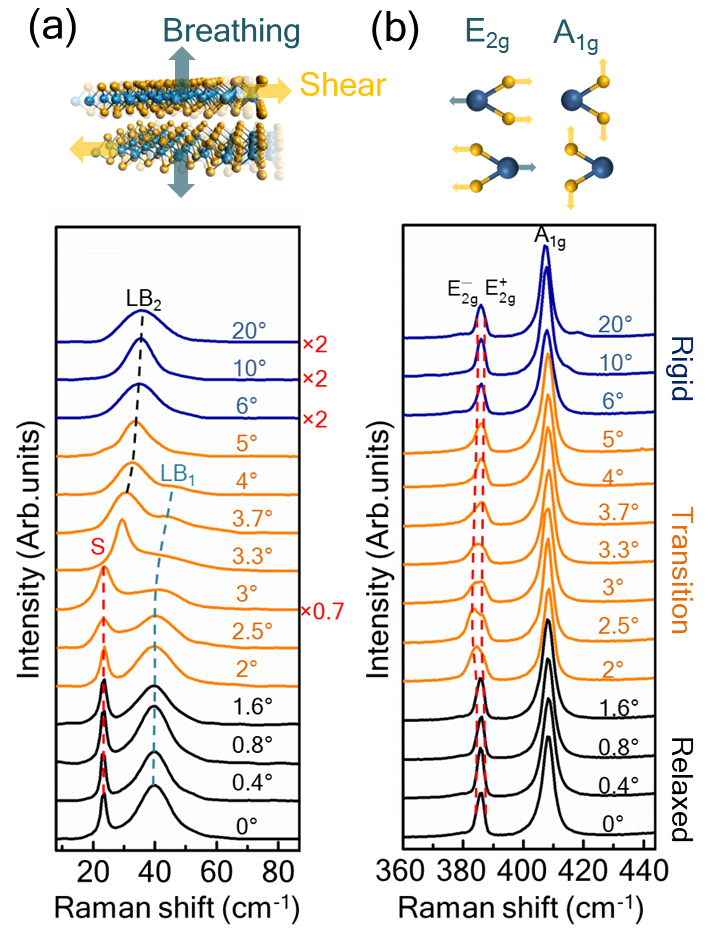}
    \caption{Measured Raman spectra of MoS$_2$ TBLs as a function of twist angle. (a) Interlayer phonon modes including the S mode and two LB (LB$_1$ and LB$_2$) modes.  (b) Intralayer phonon modes and the measured Raman spectra including the two E$_{2g}^{+}$(E$_{2g}^{-}$) and the A$_{1g}$ modes.
    The illustrations at the top show the schematic diagrams of the atomic eigenvectors for each phonon mode.
    The S mode (LB modes) corresponds to in-plane (out-of-plane) relative motions of the constituent layers, the E$_{2g}$ mode corresponds to in-plane relative motion between molybdenum and sulfur atoms and the A$_{1g}$ mode corresponds to out-of-plane sulfur atom vibrations. 
}
    \label{fig:fig2}
\end{figure}

\newpage

\begin{figure}
 \centering
   \includegraphics[width=0.8\textwidth]{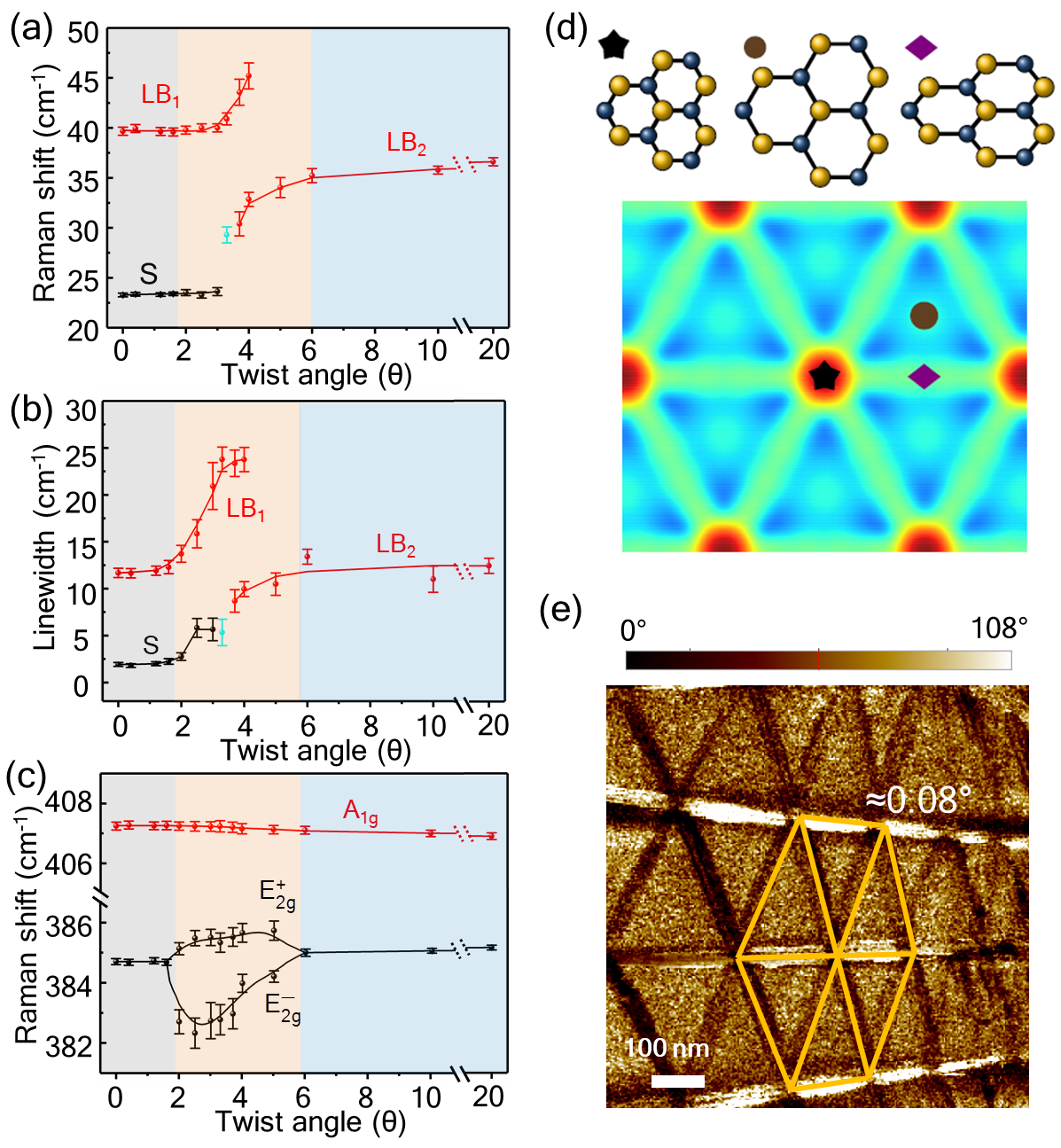}
    \caption{Analysis of the Raman spectra and experimentally observed lattice reconstruction. (a) Central frequency and (b) linewidth of S, LB$_1$ and LB$_2$ as a function of the twist angle. The cyan dots in (a) and (b) refer to a mode that cannot be uniquely identified as LB or S. (c) The central frequency of intralayer modes as a function of twist angle. \TODO{(d) Pictorial illustration of local strain at various positions within the moir\'e: compressive at the AA (black star) stacking, tensile at AB (brown circle) stacking and uni-axial along the domain boundaries (violet diamond). The corresponding positions are indicated in the strain pattern of Fig.~1d. (e) PFM phase image of reconstructed mori\'e superlattices at  $\theta\approx 0.08^{\circ}$.} 
 }
    \label{fig:fig3}
\end{figure}

\newpage

\begin{figure}
    \includegraphics[width=1.\textwidth]{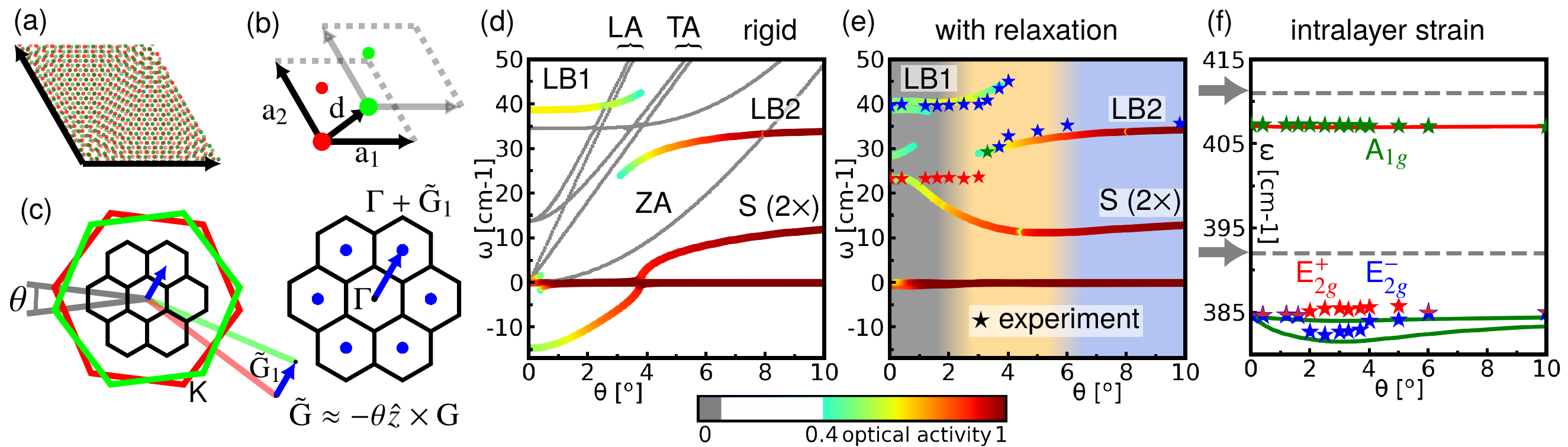}
    \caption{Calculated evolution of phonon modes as a function of twist angle $\theta$.
    (a) Real space representation of MoS$_2$ TBL at $\theta=3.9^{\circ}$. Atoms of the top (bottom) layer are illustrated in green (red). 
    (b) Locally stacked pristine unit cells.
    (c) Reciprocal representation of the moir\'e Brillouin zone (central black hexagons) and its relation to the pristine Brillouin zones (green $\&$ red).
    Blue arrows indicate relation to neighouring moir\'e $\Gamma$-points.
   Low-energy phonon evolution at $\vv q = \Gamma $, (d) neglecting or (e) including lattice relaxation as function of twist angle $\theta$. 
    Color-code is based on optical activity (see color bar) as approximated by projecting the phonon eigenmodes $\vv{Q}_{i}$ onto the central $\Gamma$-point, $|\langle \vv{Q}_{i} | \vv{\Gamma} \rangle|^{2}$. Grey lines in (d) show optically inactive modes entirely originating at
    neighboring moir\'e $\Gamma$ points (blue dots in (c)). (f) Twist-angle dependent evolution of the optical phonon modes A$_{1g}$ (red line) 
    and the degeneracy lifted E$_{2g}^{\pm}$ modes (green line). 
    Stars in (e), (f) indicate the corresponding experimental values. 
    Calculated values are shifted from gray arrows to match the experimental values at $\theta=0^{\circ}$.}
    \label{fig:fig4}
\end{figure}

\end{document}